\documentclass[aps,prl,twocolumn,superscriptaddres]{revtex4}

\usepackage{psfig}
\begin{document}
\psfigurepath{.:plot:figure}

\title{Spin dynamics in a hole-doped $S=1/2$ 
Heisenberg antiferromagnet 
with a disordered ground state}

\author{Wei Bao}
\affiliation{Los Alamos National Laboratory, Los Alamos, NM 87545}
\author{J.L. Sarrao}
\affiliation{Los Alamos National Laboratory, Los Alamos, NM 87545}

\date{\today}

\begin{abstract}
Only 3\% hole doping by Li is sufficient to suppress the
long-range antiferromagnetic order in La$_2$CuO$_4$.
Spin dynamics in such a disordered state was investigated with
measurements of the 
dynamic magnetic structure factor $S(\omega,{\bf q})$, 
using cold neutron spectroscopy, for 
La$_2$Cu$_{0.94}$Li$_{0.06}$O$_4$.
The $S(\omega,{\bf q})$ is found to sharply peak at $(\pi,\pi)$, 
and its dynamics to be relaxational. Confirming theoretical expectation
for the quantum disordered 2D $S=1/2$ Heisenberg antiferromagnet, 
the energy scale saturates at a finite value at low temperatures. 
Possible connection to the ``pseudo spin gap''
phenomenon observed in the NMR/NQR studies on underdoped cuprates is discussed.
\end{abstract}


\maketitle

Stimulated by the discovery of cuprate superconductors,
which derive from doping charge carriers into weakly 
coupled two-dimensional (2D) antiferromagnetic CuO$_2$ planes, 
there has been great interest in 2D $S$=1/2 Heisenberg antiferromagnet (HAF) 
with dominant nearest-neighbor exchange interactions on a square lattice.
It is now generally accepted that there is long-range N\'{e}el order at $T=0$
for such a system\cite{heisdm}. 
Finite-temperature magnetic 
properties\cite{la2dv}
are in good agreement with theoretical predictions\cite{2dheis}.

The situation more closely related to superconductivity in cuprates,
namely, 2D $S$=1/2 HAF with doped holes, however, is much less
understood. The parent compounds such as 
La$_2$CuO$_4$ are charge-transfer materials\cite{zsa_ct}, 
therefore, the doped hole is a charge with $S=1/2$ located at an O site. 
However, the formation of the Zhang-Rice singlet\cite{zrs} 
allows an effective description of the hole
as a spinless charge at a Cu site. Strong suppression of the N\'{e}el
order by holes can be accounted for by long-range topological
disturbances accompanying holes\cite{h_haas}. In addition,
hole motion is also disruptive to the N\'{e}el order\cite{sq2d}.
For hole-doped La$_2$CuO$_4$ without the N\'{e}el order,
two type of incommensurate antiferromagnetic correlations
have been discovered experimentally when holes introduced by dopants 
such as Sr or Ba are
mobile\cite{la2ch}. 
When holes are loosely bound in the case of Li-doped 
La$_2$CuO$_4$\cite{Li214_ryu}, antiferromagnetic correlations remain
commensurate\cite{bao99a}.
Microscopic understanding of interaction between holes and spins on the square
lattice remains a major challenge in condensed matter 
research\cite{sq2ds,stripe_jt}.
 
An alternative approach to 2D $S=1/2$ HAF with doped holes is based on
quantum phase transition. 
Building on the success of the quantum non-linear $\sigma$
model as an effective low temperature theory for 2D $S$=1/2 
HAF\cite{nlsm_wz,2dheis}, effect of doping is simulated by a
frustration parameter $g$, which at a critical value $g_c$ suppresses the 
long-range N\'{e}el order at $T=0$\cite{2dheis,2dheiqc}.
The advantage of this approach is that aspects of spin dynamics can be 
predicted from general theoretical arguments for dynamic critical
phenomena before microscopic theory is established.
The magnetically disordered state in the $g$-$T$ plane is divided into three
physical regimes: the renormalized classic, 
quantum critical (QC) and quantum disordered (QD). For doped cuprates
with disordered ground state ($g>g_c$), only the QC and QD regimes are relevant.
Quantum critical theory\cite{2dheiqc,qpt_ss}
predicts that the energy scale, $\Gamma$, is proportional to $T$ in the QC 
regime at higher $T$. Inelastic neutron scattering studies 
on Ba or Sr doped La$_2$CuO$_4$ have confirmed this prediction over an 
extraordinarily wide energy ($\hbar\omega \leq 90$ meV) and temperature 
($T\leq 500$~K) range\cite{la2smha,la2bk}.
At lower $T$ in the QD regime, $\Gamma$ is expected theoretically
to saturate at a finite value, $\Gamma_0$. 
Here we report a direct observation of such a behavior in a cold neutron 
inelastic neutron scattering study on La$_2$Cu$_{0.94}$Li$_{0.06}$O$_4$. 
With an energy resolution of 0.1 meV in term of the 
full-width-at-half-maximum of incoherent scattering, we focused on low 
energy spin dynamics, relevant in search for $\Gamma_0$.
Possible connection of the QD spin dynamics to the so-called 
``pseudo spin gap'' phenomenon observed
in the NMR/NQR studies on hole-doped La$_2$CuO$_4$ is discussed.

A single crystal of La$_2$Cu$_{0.94}$Li$_{0.06}$O$_4$,
weighing 2.1 $g$, was grown in
CuO flux, using isotopically enriched $^7$Li (98.4\%) to reduce
neutron absorption. The crystal has orthorhombic $Cmca$ symmetry with
lattice parameters $a=5.351\AA$, $b=13.15\AA$ and $c=5.386\AA$
at 295~K. Using this orthorhombic unit cell to label reciprocal {\bf q}
space, the ($\pi,\pi$) point in the square lattice notation
splits to (100) and (001) points.
Neutron scattering signal is observed
at (100) type but not at (001) type Bragg points, as in the
stoichiometric antiferromagnet La$_2$CuO$_4$\cite{la3dv}.
Measurements of seven independent (100) type of Bragg peaks, 
using thermal neutrons
to reach 6~$\AA^{-1}$, confirm the magnetic origin of these peaks.
In the remainder of this study, we will focus on
spin dynamics near {\bf Q}=(100), using cold neutron triple-axis 
spectrometer SPINS at NIST with fixed $E_f=3.7$~meV
or $E_f=5$meV. The (002) reflection of pyrolytic graphite was used for both
the monochromator and analyzer.
A cold Be or BeO filter was used to eliminate higher
order neutrons.
Horizontal Soller slits of 80' were used before and after the sample. 
Temperature of the sample was regulated by a pumped
He cryostat.

\begin{figure}
\psfig{file=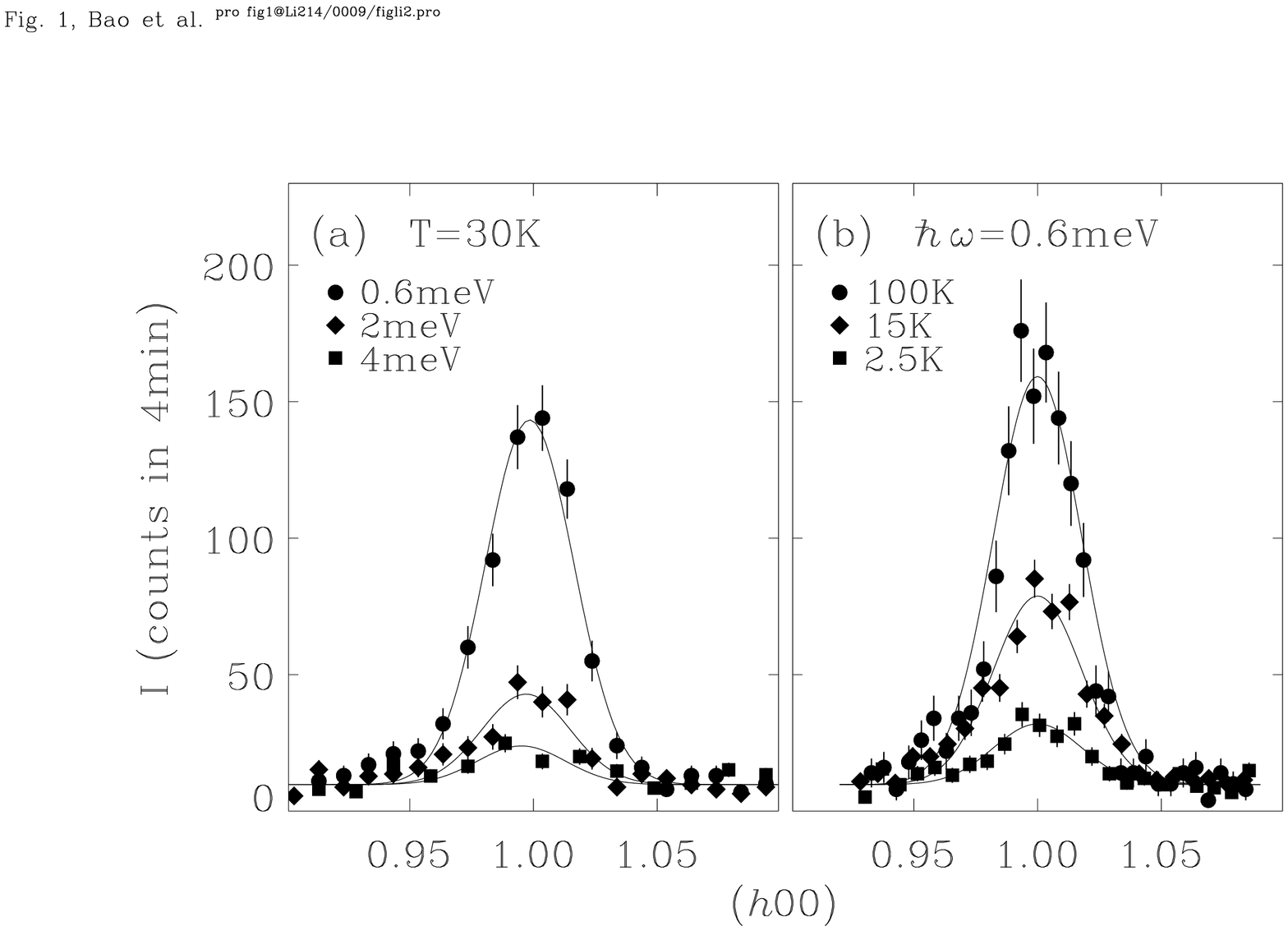,width=1.08\columnwidth,angle=90,clip=}
\vskip -10ex
\caption{
(a) Const.-$\hbar\omega$ scans with various energy transfers at 30 K.
(b) Constant-$\hbar\omega=0.6$meV scan at various temperatures. 
The {\bf q}=(100) here in the orthorhombic notation corresponds to
the ($\pi,\pi$) point of the CuO$_2$ square plane.
}
\label{fig1}
\end{figure}

Fig.~\ref{fig1} shows some constant-energy ($\hbar\omega$)
scans around {\bf Q}=(100) at various energies and temperatures,
which roughly cover the energy and temperature range of this study. 
All the peaks at ($\pi,\pi$) are resolution-limited.
This means that dynamic spin clusters with nearest-neighbor antiferromagnetic
alignment in the CuO$_2$ plane have grown to substantial size.
From the half-width-at-half-maximum of the peaks, 
a {\em lower bound} of magnetic
correlation length can be estimated, namely, $\xi > 42\AA$ below
$\sim$100~K. The correlation length is much longer than the mean
distance, 15\AA, between Li dopants. Therefore, consistent with
expectation from microscopic theory by Haas et al.\cite{h_haas}, 
antiferromagnetic correlations in our sample are not simply 
impurity limited.

\begin{figure}
\psfig{file=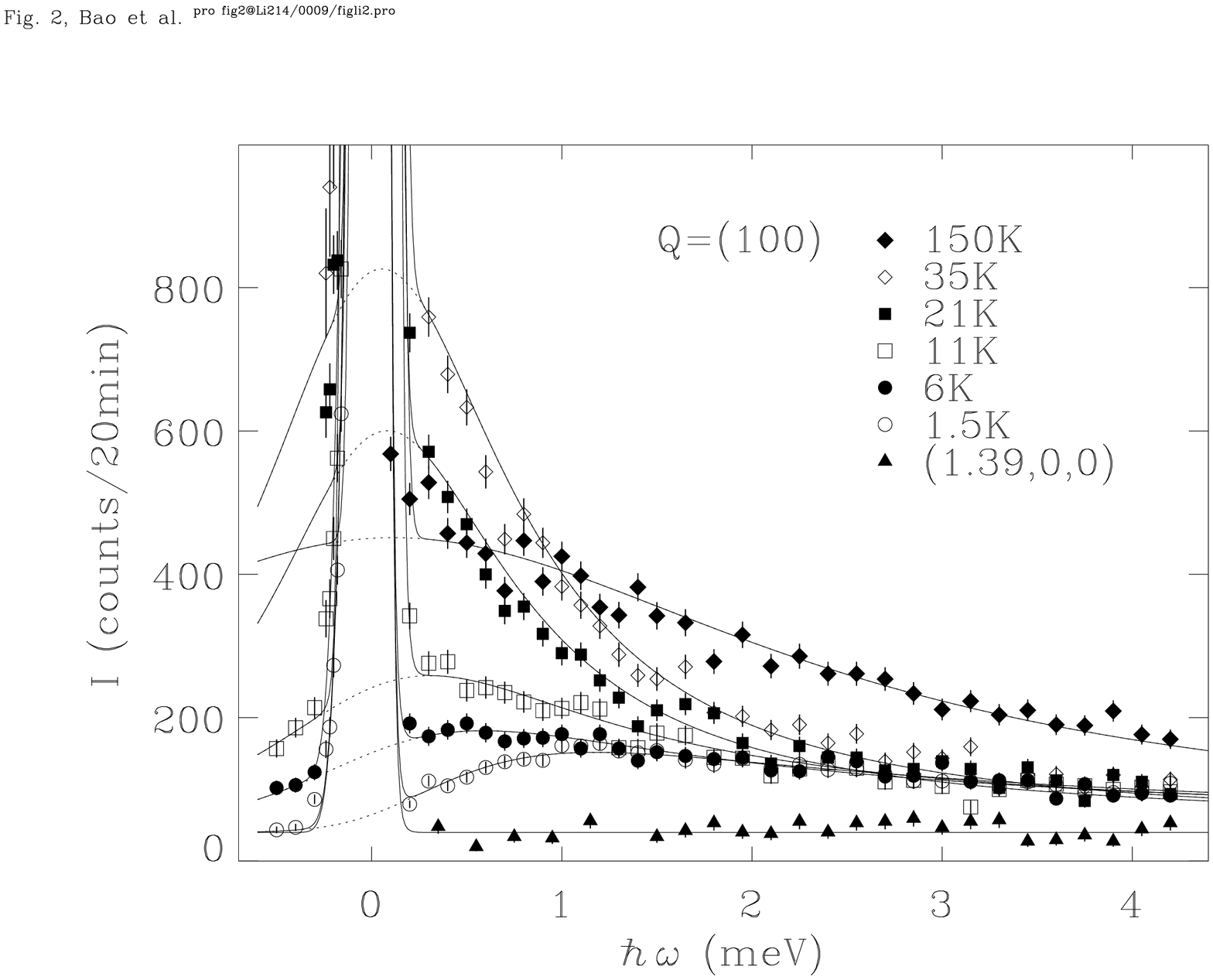,width=\columnwidth,angle=90,clip=}
\vskip -2ex
\caption{
Const.-{\bf Q}=(100) scans from 1.5 to 150~K.
The solid curves at finite $\hbar\omega$ and their dot-line continuation
near $\omega=0$ are theoretical fit to Eqs. (\ref{eq_S}) and (\ref{eq_chi})
with parameters $\Gamma$ and $\chi_{\bf Q}$ shown in Fig.~\ref{fig3}.
Triangles represent a measured flat background at (1.39,0,0).
}
\label{fig2}
\end{figure}

Dynamics of the antiferromagnetically correlated spin 
clusters in La$_2$Cu$_{0.94}$Li$_{0.06}$O$_4$ are probed by
energy scans at {\bf Q}=(100) at various temperatures (Fig.~\ref{fig2}).
Since magnetic intensity is confined within resolution at (100),
the energy scan at (1.39,0,0), which is $T$ and $\hbar\omega$-independent
(refer to triangles in Fig.~\ref{fig2}), offers a good measure of background.
Lowering temperature from 150~K, magnetic intensity at lower energies 
increases at the expense of magnetic intensity at higher energies, as shown
by filled and open diamonds. This is typical for paramagnetic fluctuations, 
as magnetic susceptibility increases while fluctuation energy decreases 
with lowering temperature\cite{crt_rev}. 
Upon further lowering temperature, however, magnetic intensity below
$\sim 1.5$meV is progressively suppressed. At 1.5~K, a peak at
a finite energy can be clearly discerned.  We would like to emphasize 
here that there is no real gap in the spin excitation spectra in 
Fig.~\ref{fig2}. As will be shown later, spectra in Fig.~\ref{fig2}
can be well described by relaxational spin dynamics.
Theoretically, both gapped and gapless spin dynamics have been found
in different toy models in the QD regime\cite{2dheiqc}.
The spectra in Fig.~\ref{fig2} also differ from inelastic neutron 
scattering spectra from
conventional, heavy fermion, and cuprate 
superconductors, which do show spin or pseudo spin gap due to superconducting 
transition\cite{scgap,ybco_rm,la2sym}.

Magnetic neutron scattering intensity is a convolution of instrument resolution 
function with dynamic magnetic structure factor $S(\omega,{\bf q})$.
Since {\bf q} scans are resolution-limited at {\bf Q}=($\pi,\pi$), 
it is convenient to write $S(\omega,{\bf q})$ as
\begin{equation}
S(\omega,{\bf q})= \frac{1}{\pi} \frac{1}{1-\exp(-\hbar\omega/k_BT)}
	\chi''_{\bf Q}(\omega) F_{\omega}({\bf q}), \label{eq_1}
\end{equation}
where $\int d{\bf q} F_{\omega}({\bf q})=1$\cite{la2smha}. Sharp structure of 
$F_{\omega}({\bf q})$ peaking at ($\pi,\pi$) can not be resolved 
in this work. Data in
Fig.~\ref{fig2} thus represent {\em local} dynamic magnetic structure factor
\begin{eqnarray}
S_{\bf Q}(\omega) & \equiv & \int d{\bf q} S(\omega,{\bf q})  \nonumber\\
 & =& \frac{1}{\pi} \frac{1}{1-\exp(-\hbar\omega/k_BT)}
	\chi''_{\bf Q}(\omega),   \label{eq_S}
\end{eqnarray}
where $\chi''_{\bf Q}(\omega)$ is the imaginary part of the {\em local} 
dynamic magnetic susceptibility.
Relaxational spin dynamics at low energy is generally described by\cite{rev_cps}
\begin{equation}
\chi''_{\bf Q}(\omega)=\frac{\hbar\omega\chi_{\bf Q}\Gamma}
	{(\hbar\omega)^2+\Gamma^2}, \label{eq_chi}
\end{equation}
where $\Gamma$ is the relaxation energy and $\chi_{\bf Q}$
represents spectral intensity.
For scan at each temperature in Fig.~\ref{fig2}, $\Gamma$ and
$\chi_{\bf Q}$ are extracted by least squared fit
of the data to Eqs (\ref{eq_S}) and (\ref{eq_chi}), plus the usual 
elastic/incoherent peak, convoluted with instrument resolution. 
A flat background measured at {\bf q}=(1.39,0,0) (refer to triangle)
is included in the fitting. The consistency between the theoretical curves
and measured data in Fig.~\ref{fig2} is very satisfactory.

\begin{figure}
\centerline{\psfig{file=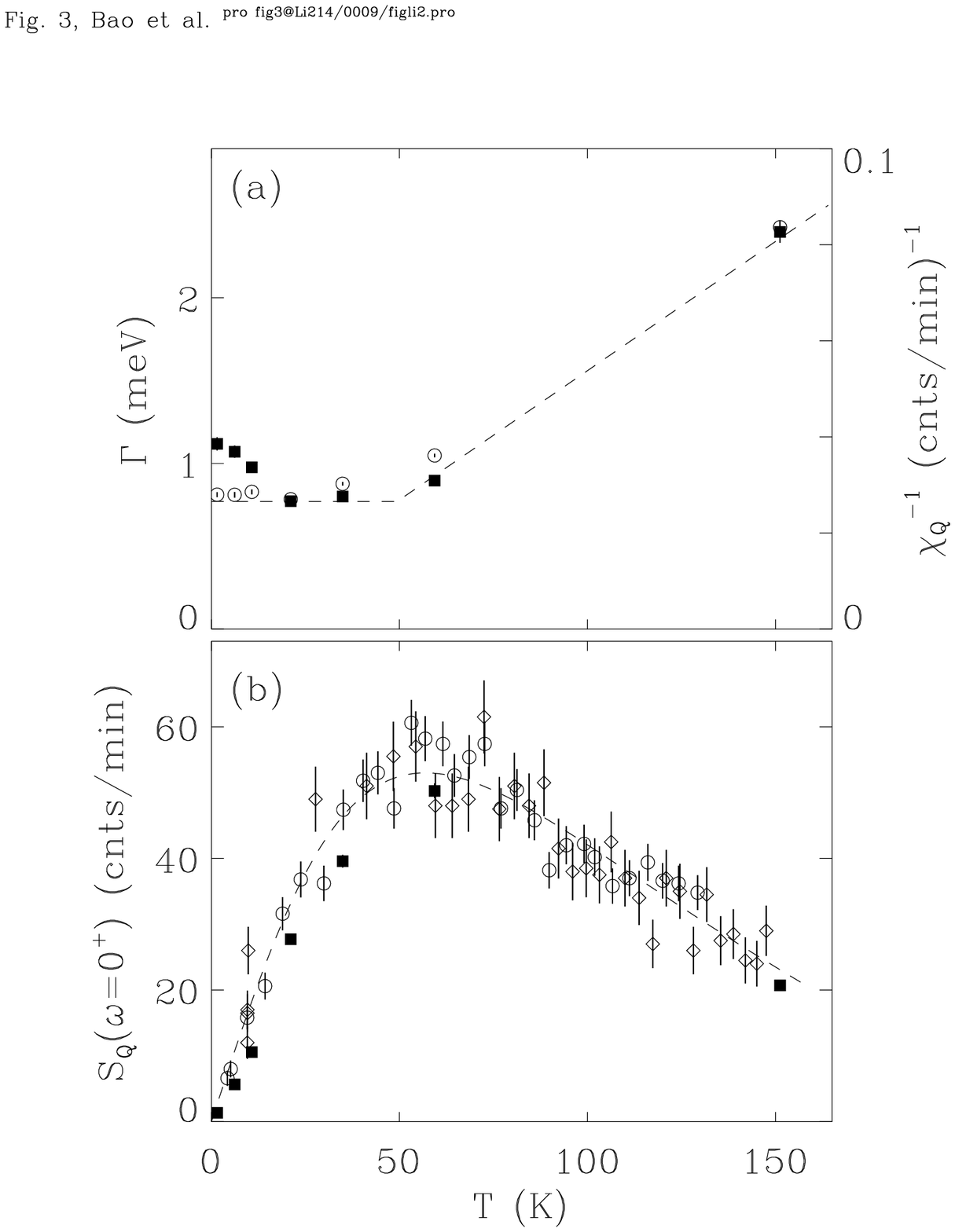,width=.7\columnwidth,angle=0,clip=}}
\vskip -4ex
\caption{\label{fig3}
(a) Measured temperature dependence of $\Gamma$ (squares)
and $\chi_{\bf Q}^{-1}$ (circles) for 
La$_2$Cu$_{0.94}$Li$_{0.06}$O$_4$ [Refer to Eq.\ (\ref{eq_chi})].
(b) Temperature dependence of quasielastic magnetic intensity,
$S_{\bf Q}(\omega=0^+)$. 
The filled squares are calculated using Eq.~(\ref{eq_5}) from $\Gamma$
and $\chi_{\bf Q}$ in (a), and open circles and diamonds were measured
at low energy, $\hbar\omega=0.2$~meV, during warming and cooling, respectively.
}
\end{figure}

Experimental $\Gamma$ and $\chi_{\bf Q}^{-1}$ as a function of temperature
are shown in Fig.~\ref{fig3}(a) by square and circle, respectively.
Clearly, $\Gamma$ and $\chi_{\bf Q}^{-1}$ saturate at finite values 
at low temperatures.
It is well known for conventional magnetic materials 
that in paramagnetic phase, $\Gamma$
is a monotonically increasing function of $T$ which is zero at
magnetic transition temperature\cite{crt_rev}. This behavior applies also to
spin-freezing transition of classic spin glasses such as $Cu$-Mn\cite{sglass}. 
When magnetic transition is prevented
by frustration, the monotonic behavior for $\Gamma$ still persists\cite{shlee}.
Temperature independent $\Gamma$ has been observed in intermediate-valence
compounds\cite{iv_ehm}. However, there is no magnetic correlations between
rare-earth ions in these materials and the $\Gamma$ reflects valence
fluctuations. In La$_2$Cu$_{0.94}$Li$_{0.06}$O$_4$,
magnetic correlations are substantial with $\xi >42\AA$. What we observe
here is consistent with theoretical expectation for 2D $S$=1/2 HAF
with the QD ground state\cite{qpt_ss}:
$\Gamma(T)\sim \max[k_B T,\, (g-g_c)^{z\nu}]$. 
The dashed line in Fig.~\ref{fig3}(a),
$\Gamma(T)=\max[0.18 k_B T,\, \Gamma_0]$ 
with $\Gamma_0=0.77$meV, captures the main feature of our data. 

In hole-doped La$_2$CuO$_4$, the $g$ is physically the doping concentration.
Besides frustrating the long-range antiferromagnetic order, doping inevitably
introduces disorders. The spin freezing temperature $T_{sf}$ measured with
$\mu$SR for Li-doped La$_2$CuO$_4$\cite{Li214muSR} is nearly identical to that
for Sr-doped La$_2$CuO$_4$\cite{fchou}. 
For La$_2$Cu$_{0.94}$Li$_{0.06}$O$_4$, $T_{sf}=8$~K. Up to now, 
little has been 
predicted for $S(\omega,{\bf q})$ for realistic spin glasses. In 
ferromagnetic reentrant spin glasses, spin dynamics expected of a ferromagnet
has been experimentally observed within a finite range of {\bf q} of magnetic
zone center in the temperature interval $T_{sf} <T<T_C$\cite{FeAl_msm}.
Along this line, one would expect the QD behavior to hold between 
$\sim$10 and $\sim$60~K for La$_2$Cu$_{0.94}$Li$_{0.06}$O$_4$, refer to Fig.~\ref{fig3}(a). 
In addition, below spin freezing, $\Gamma$ from spin clusters at ($\pi,\pi$) in 
La$_2$Cu$_{0.94}$Li$_{0.06}$O$_4$ remains
finite. There is no sign of freezing $\Gamma \rightarrow 0$, which is 
a universal feature for classic spin glasses. 
This suggests that inelastic magnetic intensity in Fig.~\ref{fig2}
comes from spin clusters which do not participate in spin freezing.
The upturn of $\Gamma$ in Fig.~\ref{fig3}(a) below 11~K may be attributed to
modification to the QD dynamics due to coupling of the fluctuating
spin clusters to freezing spin clusters.

One motivation of this low-energy neutron scattering work is to make 
connection to Cu nuclear resonance NMR/NQR results. 
Soon after the discovery of cuprate superconductivity,
anomalous suppression of low energy spin fluctuations was discovered in
{\em normal} state of various family of underdoped 
cuprates\cite{takigawa,sfuji}.  
Refer to Ref.~\cite{rev_cps} for a historic review of this so-called
``pseudo spin-gap'' phenomenon (PSG).
It has been discussed variously as due to, e.g., a temperature
dependent magnetic correlation length\cite{tmnmr,pinesnmr}, 
resonating valence bond singlet pairing\cite{rvbt}, 
activation gaps of stripe glasses\cite{pavia,Eunjc}, or
incipient superconducting order\cite{ybco_rm}. In the
La$_2$CuO$_4$ family, extensive inelastic neutron scattering studies
have found no such gap\cite{la2smha,la2bk}, 
except the spin gap due to superconducting
transition\cite{la2sym}.

The nuclear spin-lattice relaxation rate $1/T_1$ is a weighted summation
of dynamic magnetic structure factor $S(\omega,{\bf q})$ over the 
Brillouin zone at small $\omega=0^+$,
\begin{equation}
T_1^{-1}=\sum_{\bf q} \left|A({\bf q})\right|^2 S(0^+,{\bf q}),
\end{equation}
where hyperfine coupling $\left|A({\bf q})\right|^2$ peaks at ($\pi,\pi$)
for Cu NMR/NQR\cite{rev_cps}.
Using Eqs. (\ref{eq_1})-(\ref{eq_chi}) for $S(\omega,{\bf q})$,
\begin{eqnarray}
T_1^{-1}&=& S_{\bf Q}(0^+) 
	\sum_{\bf q} \left|A({\bf q})\right|^2 F_{\omega}({\bf q}) \nonumber\\
	&\sim &S_{\bf Q}(0)=\frac{\chi_{\bf Q}k_BT}{\pi \Gamma}. \label{eq_5}
\end{eqnarray}
In Fig.~\ref{fig3}(b), $S_{\bf Q}(0)$ calculated using Eq.~(\ref{eq_5})
from the experimentally determined $\chi_{\bf Q}$ and $\Gamma$ in Fig.~\ref{fig3}(a) are
shown together with $S_{\bf Q}(\omega)$ measured at $\hbar\omega=0.2$~meV,
which is the lowest energy transfer without significant
elastic contamination (refer to Fig.~\ref{fig2}). 
Circles (diamonds) were measured during warming (cooling) cycle. 
These data, $S_{\bf Q}(\omega\sim 0)\sim 1/T_1$, 
bear remarkable resemblance to $1/T_1$ observed in Sr-doped La$_2$CuO$_4$ 
of similar hole concentration\cite{sfuji}. 
The reduction of $1/T_1$ at low $T$ below the extrapolation from high $T$
behavior is the experimental observation termed ``pseudo spin gap''
in NMR/NQR studies on cuprates. 
We have shown that there is no gap in spin excitation
spectra for La$_2$Cu$_{0.94}$Li$_{0.06}$O$_4$. The apparent PSG behavior in
Fig.~\ref{fig3}(b) is merely a consequence of a saturating $\Gamma$ at low $T$,
which is expected for 2D $S$=1/2 HAF of $g>g_c$ with the QD ground state.
Thus a possible link between the NMR/NQR PSG behavior with the QD state
is suggested by this study; specifically, 
the PSG phenomenon may not require a pseudo
gap for its explanation.

From this point of view, if we take doped cuprates
without the N\'{e}el order as the experimental realization of $g>g_c$, 
then the ubiquitous NMR/NQR PSG behavior is expected as a consequence of a finite
$\Gamma_0\sim (g-g_c)^{z\nu}$. 
The absence of spin or pseudo spin gap from inelastic neutron scattering 
studies can then also be reconciled with the PSG from NMR/NQR studies.
Similar NMR/NQR spectra for underdoped La$_2$CuO$_4$\cite{pavia,Eunjc,awhti,fchou}
may also be understood as a consequence that the
dopant elements, dopant location (in or out-of-plane) and crystal structure
(orthorhombic or tetragonal) are irrelevant perturbations to the QD regime.
As such, study of the PSG phenomenon should
pay at least as much attention to saturating $\Gamma$ as to the
presence of a real spin gap.

In summary, we found dynamic spin clusters in hole-doped La$_2$Cu$_{0.94}$Li$_{0.06}$O$_4$ to remain ($\pi,\pi$)-correlated. 
Spin dynamics of such clusters, with correlation length much larger than
mean impurity distance below 150~K, is relaxational at ($\pi,\pi$).
Confirming predictions of quantum critical theory for 2D
$S=1/2$ Heisenberg antiferromagnet in the quantum disordered regime, 
the energy scale $\Gamma$
of La$_2$Cu$_{0.94}$Li$_{0.06}$O$_4$ saturates around 0.8 meV below $\sim$60~K.
We discuss the possible connection between the so-called ``pseudo spin-gap'' 
phenomenon discovered in NMR/NQR studies on underdoped cuprates
and the ``quantum disordered'' behavior observed here. In other words,
the NMR/NQR ``pseudo spin-gap'' phenomenon may be accounted for 
without pseudo spin gap in spin excitation spectra.

We thank L. Yu, E. Tosatti, C.M. Varma, S. Sachdev,
G. Aeppli, J. Haase, P. Carretta, P.C. Hammel, N.J. Curro, 
E. Dagotto, G. Kotliar, A.V. Balatsky, Y. Bang, A. Abanov, D. Pines,
R. Heffner, and A.P. Ramirez for useful discussions; 
we thank S.-H. Lee for hospitality and assistance at NIST. 
SPINS at NIST is supported by NSF. Work at LANL is
supported by U.S. Dept.\ of Energy.


\begin{thebibliography}{77}
\expandafter\ifx\csname natexlab\endcsname\relax\def\natexlab#1{#1}\fi
\expandafter\ifx\csname bibnamefont\endcsname\relax
  \def\bibnamefont#1{#1}\fi
\expandafter\ifx\csname bibfnamefont\endcsname\relax
  \def\bibfnamefont#1{#1}\fi
\expandafter\ifx\csname citenamefont\endcsname\relax
  \def\citenamefont#1{#1}\fi
\expandafter\ifx\csname url\endcsname\relax
  \def\url#1{\texttt{#1}}\fi
\expandafter\ifx\csname urlprefix\endcsname\relax\def\urlprefix{URL }\fi
\providecommand{\bibinfo}[2]{#2}
\providecommand{\eprint}[2][]{\url{#2}}

\bibitem{heisdm}
{ D. C. Mattis and C. Y. Pan, Phys. Rev. Lett.
  {\bf 61}, 463 (1988);
R. R. P. Singh, Phys. Rev. B {\bf 39}, 9760
  (1989);
J. Igarashi, {\it ibid.} {\bf 46}, 10763
  (1992)}.

\bibitem{la2dv}
{G. Shirane et al., Phys. Rev. Lett. \textbf{59},
  1613 (1987);
M. Greven et al., {\it ibid.} \textbf{7},
  1096 (1994);
H.~M. R{\o}nnow et~al., {\it ibid.}
  \textbf{82}, 3152 (1999);
  \textbf{87}, 037202 (2001);
K.~Yamada et~al.,
  Phys. Rev. B \textbf{40},
  4557 (1989);
R.~J. Birgeneau et~al., {\it ibid.} \textbf{59}, 13788 (1999)}.

\bibitem{2dheis}
{ S. Chakravarty et al., Phys. Rev. Lett. {\bf 60}, 1057 (1988);
A. Auerbach et al., {\it ibid.} {\bf
  61}, 617 (1988);
A.~Cuccoli et~al., {\it ibid.} \textbf{77},
  3439 (1996);
B.~B. Beard et~al., {\it ibid.}
  \textbf{80}, 1742 (1998);
J.-K. Kim et~al.,
  {\it ibid.} \textbf{80},
  2705 (1998);
M.~S. Makivic et~al., Phys. Rev. B  \textbf{43}, 3562 (1991)}.

\bibitem{zsa_ct}
{{J. Zaanen et al., Phys. Rev. Lett. {\bf 55}, 418 (1985)}.

\bibitem[{\citenamefont{Zhang and Rice}(1988)}]{zrs}
\bibinfo{author}{\bibfnamefont{F.~C.} \bibnamefont{Zhang}} \bibnamefont{and}
  \bibinfo{author}{\bibfnamefont{T.~M.} \bibnamefont{Rice}},
  \bibinfo{journal}{Phys. Rev. B} \textbf{\bibinfo{volume}{37}},
  \bibinfo{pages}{3759} (\bibinfo{year}{1988}).

\bibitem{h_haas}
{S.~Haas et~al., Phys. Rev. Lett. \textbf{77},
  3021 (1996);
C.~Timm and K.~H. Bennemann,
  {\it ibid.} \textbf{84}, 4994 (2000)}.

\bibitem{sq2d}
{ B.~I.\ Shraiman and E.~D.\ Siggia, Phys. Rev.
  Lett. {\bf 61}, 467 (1988)}.

\bibitem{la2ch}
{S.-W. Cheong et al., Phys. Rev. Lett.
  \textbf{67}, 1791 (1991);
K.~Yamada et~al., Phys. Rev. B \textbf{57}, 6165 (1998);
M.~Matsuda et~al.,
  {\it ibid.} \textbf{62},
  9148 (2000);
H.~A. Mook et~al.,
  Nature \textbf{395}, 580
  (1998)}.

\bibitem{Li214_ryu}
{A.~I. Rykov et al., Physica C \textbf{247},
  327 (1995);
J.~L. Sarrao et al., Phys. Rev. B
  \textbf{54}, 12014 (1996)}.

\bibitem{bao99a}
{W.~Bao et al., Phys. Rev. Lett. \textbf{84},
  3978 (2000)}.

\bibitem{sq2ds}
{ B. I. Shraiman and E. D. Siggia, Phys. Rev.
  Lett. {\bf 62}, 1564 (1989);
H. J. Schulz, Phys. Rev. Lett. {\bf 64}, 1445
  (1990);
S. Sarker et al., Phys. Rev. B {\bf 43}, 8775 (1991);
F. Yuan et~al. {\it ibid.} \textbf{64},
  224505 (2001)}.

\bibitem{stripe_jt}
{J.~M. Tranquada et al., Nature \textbf{375}, 561 (1995); 
V.~J. Emery et al., Phys. Rev. Lett. \textbf{64}, 475 (1990); 
A.~L. Chernyshev et~al., {\it ibid.} \textbf{84}, 4922 (2000);
E.~Dagotto et~al., Phys. Rev. B \textbf{49}, 3548 (1994);
T.~Tohyama et~al., {\it ibid.} \textbf{59}, R11649 (1999);
 T. Giamarchi and C. Lhuillier, {\it ibid.} {\bf 42}, 10641 (1990);
C.~Buhler et~al., {\it ibid.} \textbf{62}, R3620 (2000)}.

\bibitem{nlsm_wz}
{X.~G. Wen and A.~ Zee, Phys. Rev. Lett. \textbf{ 61}, 1025 (1988);
F.~D.~M. Haldane, {\it ibid.} \textbf{61}, 1029 (1988);
E.~Fradkin and M.~Stone, Phys. Rev. B \textbf{38}, 7215 (1988)}.

\bibitem{2dheiqc}
{ S. Sachdev and J. Ye, Phys. Rev. Lett. {\bf 69}, 2411 (1992); {\bf
  70}, 3339 (1993); {\bf 70}, 4011 (1993)}.

\bibitem{qpt_ss}
\bibinfo{author}{\bibfnamefont{S.}~\bibnamefont{Sachdev}},
  \emph{\bibinfo{title}{Quantum Phase Transitions}}
  (\bibinfo{publisher}{Cambridge University Press},
  \bibinfo{address}{Cambridge}, \bibinfo{year}{1999}).

\bibitem{la2smha}
{S.~M. Hayden et al., Phys. Rev. Lett. \textbf{66}, 821 (1991)}.

\bibitem{la2bk}
{B. Keimer et~al. Phys. Rev. Lett. \textbf{ 67}, 1930 (1991)}.

\bibitem{la3dv}
{D.~Vaknin et al., Phys. Rev. Lett. \textbf{58}, 2802 (1987)}.

\bibitem[{\citenamefont{Hohenberg and Halperin}(1977)}]{crt_rev}
\bibinfo{author}{\bibfnamefont{P.~C.} \bibnamefont{Hohenberg}}
  \bibnamefont{and} \bibinfo{author}{\bibfnamefont{B.}~\bibnamefont{Halperin}},
  \bibinfo{journal}{Rev. Modern Phys.} \textbf{\bibinfo{volume}{49}}
  \bibinfo{pages}{435} (\bibinfo{year}{1977}).

\bibitem{scgap}
{C.~G. Shull and F.~A. Wedgwood, Phys. Rev. Lett. \textbf{16}, 513 (1966);
N.~Metoki et~al., {\it ibid.} \textbf{80}, 5417 (1998)}.

\bibitem{ybco_rm}
{ J. Rossat-Mignod et~al., Physica B \textbf{169} 58 (1993)}.

\bibitem{la2sym}
{T.~E. Mason et al., Phys. Rev. Lett. \textbf{71}, 919 (1993)}.

\bibitem[{\citenamefont{Slichter}(1994)}]{rev_cps}
\bibinfo{author}{\bibfnamefont{C.~P.} \bibnamefont{Slichter}}, in
  \emph{\bibinfo{booktitle}{Strongly Correlated Electronic Materials}}, edited
  by \bibinfo{editor}{\bibfnamefont{K.~S.} \bibnamefont{Bedell et al.}}
  (\bibinfo{publisher}{Addison-Wesley Publishing}, \bibinfo{address}{Reading},
  \bibinfo{year}{1994}), p. \bibinfo{pages}{427}.

\bibitem[{\citenamefont{Mezei and Murani}(1979)}]{sglass}
\bibinfo{author}{\bibfnamefont{F.}~\bibnamefont{Mezei}} \bibnamefont{and}
  \bibinfo{author}{\bibfnamefont{A.~P.} \bibnamefont{Murani}},
  \bibinfo{journal}{J. Magn. Magn. Mater.} \textbf{\bibinfo{volume}{14}},
  \bibinfo{pages}{211} (\bibinfo{year}{1979}).

\bibitem[{\citenamefont{Lee et~al.}(2001)}]{shlee}
\bibinfo{author}{\bibfnamefont{S.-H.} \bibnamefont{Lee}} \bibnamefont{et~al.},
  \bibinfo{journal}{Phys. Rev. Lett.} \textbf{\bibinfo{volume}{86}},
  \bibinfo{pages}{5554} (\bibinfo{year}{2001}).

\bibitem{iv_ehm}
{E.~ Holland-Moritz et~al., Phys. Rev. B \textbf{25}, 7482 (1982)}.

\bibitem{Li214muSR}
{L.~P. Le et al., Phys. Rev. B \textbf{54}, 9538 (1996)}.

\bibitem{fchou}
{F.~C. Chou et al., Phys. Rev. Lett. \textbf{71}, 2323 (1993); 
C. Niedermayer et~al., {\it ibid.} \textbf{80}, 3843 (1998)}.

\bibitem{FeAl_msm}
{K.~Motoya et~al., Phys. Rev. B \textbf{28}, 6183 (1983);
W. Bao et~al., Phys. Rev. Lett. \textbf{82},
  4711 (1999)}.

\bibitem{takigawa}
{M.~ Takigawa et~al., Phys. Rev. B \textbf{43}, 247 (1991); 
T. Imai et al., Phys. Rev. Lett. {\bf 70}, 1002 (1993)}.


\bibitem{sfuji}
{S. Fujiyama et~al., J. Phys. Soc. Jpn. \textbf{66}, 2864 (1997);
S.~Ohsugi et~al., {\it ibid.} \textbf{63}, 700 (1994);
A. Goto et~al., {\it ibid.} \textbf{65}, 3401 (1996);
Y. Itoh et~al. {\it ibid.} \textbf{65}, 3751 (1996)}.

\bibitem{tmnmr}
{T.~Moriya et al., J. Phys. Soc. Jpn. \textbf{59},
  2905 (1990)}.

\bibitem[{\citenamefont{Barzykin and Pines}(1995)}]{pinesnmr}
\bibinfo{author}{\bibfnamefont{V.}~\bibnamefont{Barzykin}} \bibnamefont{and}
  \bibinfo{author}{\bibfnamefont{D.}~\bibnamefont{Pines}},
  \bibinfo{journal}{Phys. Rev. B} \textbf{\bibinfo{volume}{52}},
  \bibinfo{pages}{13585} (\bibinfo{year}{1995}).

\bibitem{rvbt}
{T.~Tanamoto et al., J. Phys. Soc. Jpn. \textbf{63},
  2739 (1994)}.

\bibitem[{\citenamefont{Julien et~al.}(1999)}]{pavia}
\bibinfo{author}{\bibfnamefont{M.-H.} \bibnamefont{Julien}}
  \bibnamefont{et~al.}, \bibinfo{journal}{Phys. Rev. Lett.}
  \textbf{\bibinfo{volume}{83}}, \bibinfo{pages}{604} (\bibinfo{year}{1999}).

\bibitem[{\citenamefont{Curro et~al.}(2000)}]{Eunjc}
\bibinfo{author}{\bibfnamefont{N.~J.} \bibnamefont{Curro}}
  \bibnamefont{et~al.}, \bibinfo{journal}{Phys. Rev. Lett.}
  \textbf{\bibinfo{volume}{85}}, \bibinfo{pages}{642} (\bibinfo{year}{2000}).

\bibitem{awhti}
{A.~W. Hunt et al., Phys. Rev. Lett. \textbf{82}, 4300 (1999); 
B.~J. Suh et~al., {\it ibid.} \textbf{81}, 2791 (1998)}.

}
\end{thebibliography}
\end{document}